\documentclass[fleqn,usenatbib]{mnras}
\usepackage{newtxtext,newtxmath}
\usepackage[T1]{fontenc}
\usepackage{graphicx}
\usepackage{amsmath}
\usepackage{amssymb}

\newcommand{\msun}{{\rm M}_\odot}
\newcommand{\nh}{n_{\rm H}}
\newcommand{\cc}{{\rm cm^{-3}}}

\title[First star formation in FDM cosmology]{First star formation in ultra-light particle dark matter cosmology}

\author[Hirano et al.]{
Shingo Hirano,$^{1}$\thanks{E-mail: shirano@astro.as.utexas.edu}
James M. Sullivan$^{1}$
and
Volker Bromm$^{1}$
\\
$^{1}$Department of Astronomy, University of Texas, Austin, TX 78712, USA
}

\date{Accepted 2017 September 19. Received 2017 September 19; in original form 2017 May 25}
\pubyear{2017}

\begin{document}
\label{firstpage}
\pagerange{\pageref{firstpage}--\pageref{lastpage}}
\maketitle

\begin{abstract}

The formation of the first stars in the high-redshift Universe is a sensitive probe of the small-scale, particle physics nature of dark matter (DM).
We carry out cosmological simulations of primordial star formation in ultra-light, axion-like particle DM cosmology, with masses of $10^{-22}$ and $10^{-21}$\,eV, with de Broglie wavelengths approaching galactic scales ($\sim$~kpc).
The onset of star formation is delayed, and shifted to more massive host structures.
For the lightest DM particle mass explored here, first stars form at $z \sim 7$ in structures with $\sim\!\! 10^9\,\msun$, compared to the standard minihalo environment within the $\Lambda$ cold dark matter ($\Lambda$CDM) cosmology, where $z \sim 20$ -- $30$ and $\sim\!\!10^5$ -- $10^6\,\msun$.
Despite this greatly altered DM host environment, the thermodynamic behaviour of the metal-free gas as it collapses into the DM potential well asymptotically approaches a very similar evolutionary track.
Thus, the fragmentation properties are predicted to remain the same as in $\Lambda$CDM cosmology, implying a similar mass scale for the first stars.
These results predict intense starbursts in the axion cosmologies, which may be amenable to observations with the {\it James Webb Space Telescope}.

\end{abstract}


\section{Introduction}
\label{sec:intro}

The particle physics nature of the dynamically dominant dark matter component of the cosmic energy density is one of the crucial open questions in modern science.
Whereas the $\Lambda$ cold dark matter ($\Lambda$CDM) model successfully explains the observed phenomenology
on scales larger than galaxies, its extrapolation to smaller scales is largely untested.
Until recently, supersymmetric weakly interacting massive particles (WIMPs) were considered the most promising candidates for dark matter, due to a number of extremely appealing, seemingly `natural' aspects, such as the WIMP miracle \citep[e.g.][]{bertone10}.
However, in light of the failed experimental efforts to date to detect such WIMPs in the laboratory, alternative scenarios for the nature of dark matter are explored with renewed vigour.
In guiding this theoretical exploration, astrophysical constraints are of key importance.
The primordial power spectrum has been directly probed only down to length scales of $\simeq\!30$\,comoving\,Mpc \citep{hlozek12}, and we need to resort to non-linear probes to constrain the structure formation model at even smaller scales.
Specifically, it has been realized for a while that high-redshift structure formation, and in particular the formation of the first stars, the so-called Population~III (Pop~III), provides a sensitive probe of the small-scale nature of dark matter \citep[e.g.][]{yoshida03,dayal17}.

Pop~III star formation has been investigated in great detail within standard $\Lambda$CDM, where minihaloes of virial mass $\sim\!\!10^{5}$ -- $10^{6}\,\msun$ at $z \sim 20$ -- $30$ are identified as host sites, resulting in a top-heavy distribution of stellar masses \citep[e.g.][]{bromm13,greif15}.
It is an open question how the character of Pop~III star formation will change with variations of the underlying dark matter cosmology.
As an example, \cite{hirano15b} studied one possible effect by considering a primordial power spectrum with a blue tilt, enhancing power on small scales, leading to a dramatic transformation of the first star formation process.

{\it What, then, are possible alternatives to WIMP dark matter?}
It is well known that $\Lambda$CDM is challenged on small scales by a number of discrepancies between prediction and observation, such as the `missing satellites' problem \citep[e.g.][]{bullock13}, the `core-cusp' problem \citep[e.g.][]{deblok10}, and the `too big to fail' problem \citep[e.g.][]{boylan-kolchin12}.
One class of proposed solutions invokes baryonic physics, such as star formation, supernovae, and black hole feedback, acting to decouple the luminous component of galaxies from their dark matter, or dynamically transforming their structure \citep[e.g.][]{wetzel16}.
On the other hand, the nature of DM itself can alleviate those problems, postulating less massive particles whose intrinsic motions can quench perturbation growth and structure formation below their free-streaming scale.
An intriguing theoretical model involves ultra-light particles, like axions \citep[e.g.][and references therein]{marsh16a}, with masses about $10^{-22}$\,eV.
Their corresponding de Broglie wavelength is macroscopically large, $\approx\!\!1$\,kpc, effectively suppressing growth of small-scale structure \citep[see][for a recent review]{hui17}.

This DM candidate is sometimes termed {\it fuzzy dark matter} (FDM), because of the quantum mechanical origin of establishing stability, similar to that in atoms and molecules \citep{hu00}.
In such ultra-light DM particle universes, non-linear structure formation can be prevented on small scales, with the suppression (Jeans) scale given by the de Broglie wavelength.
This opens up the possibility to naturally solve the small-scale problems within the $\Lambda$CDM model, thus accommodating most observations,
e.g. Lyman-$\alpha$ forest constraints \citep{irsic17,zhang17}.
In particular, any such suppression of small-scale structure can significantly affect the formation of the first objects in the Universe \citep[e.g.][]{greif15}.
Due to a lack of numerical simulations in a fully cosmological context, however, the exact impact of the FDM model on the early Universe has not yet been established \citep[e.g.][and their upcoming paper II]{mocz17toy}.
Thus, remarkably the Heisenberg uncertainty relation may be manifest on galactic scales, stabilizing the dark matter against gravitational instability on these scales.
Such a `soliton' solution has been explored in-depth in the context of the core-cusp problem, and is approximately consistent with an ultra-light axion mass range of $\approx\!10^{-22}$ -- $10^{-21}$\,eV \citep{marsh15}.

In this {\it Letter}, we investigate first object formation in such ultra-light dark matter cosmologies by performing a series of hydrodynamic simulations.
The suppression of small-scale structure dramatically alters the nature of the first star and galaxy formation sites, rendering the DM host structures much more massive, with correspondingly higher virial temperatures.
One consequence is that the gaseous fuel available for Pop~III star formation may be greatly enhanced, thus possibly triggering luminous primordial starbursts that in turn could be within reach of the {\it James Webb Space Telescope (JWST)}.
The cooling of the metal-free gas, however, still has to rely on molecular hydrogen (H$_2$), and it is an important question whether the characteristic thermodynamics that imprints a fragmentation scale of a few $\sim\!\!100\,\msun$ in the case of standard $\Lambda$CDM cosmology remains intact, or is critically different \citep[e.g.][]{bromm13}.

The remainder of the paper is organized as follows.
We begin by describing our numerical methodology in Section~\ref{sec:method}.
Sections~\ref{sec:halo} and \ref{sec:cloud} present simulation results, first regarding the DM host environment and then the physical properties of the collapsing cloud.
Sections~\ref{sec:summary} summarizes the implications of adopting FDM cosmologies.
Throughout this study, we adopt cosmological parameters with matter density $\Omega_{\rm m} = 0.31$, baryon density $\Omega_{\rm b} = 0.048$, dark energy density $\Omega_{\Lambda} = 0.69$, a Hubble constant of $h = 0.68$, a normalization of the density fluctuations of $\sigma_8 = 0.83$, and a primordial power spectral index of $n_{\rm s} = 0.96$ \citep{PLANCK13XVI}.

\section{Numerical Methodology}
\label{sec:method}

We perform a series of cosmological simulations started from initial conditions representing different FDM models.
The cosmological initial conditions are generated using the publicly available code {\sc music} \citep{hahn11}.
The linear power spectrum $P_{\rm FDM}(k)$ for FDM cosmology, with the mass of the ultra-light dark matter particle being $m_{\rm a}$, is given by assuming a transfer function $T_{\rm FDM}(k)$ \citep{hu00} as
\begin{eqnarray}
&&P_{\rm FDM}(k) = T_{\rm FDM}^2(k) P_{\rm CDM}(k) \, , \\
&&\begin{cases}
T_{\rm FDM}(k) = \cos x_{\rm J}^3(k) / (1 + x_{\rm J}^8(k)) \, , \\
x_{\rm J}(k) \ \ \ \ \ \, = 1.61 ( m_{\rm a} / 10^{-22}\,{\rm eV}/c^2 )^{1/18} ( k / k_{\rm J,eq} ) \, , \\
k_{\rm J,eq} \ \ \ \ \ \ \, = 9 ( m_{\rm a} / 10^{-22}\,{\rm eV}/c^2 )^{1/2}\,{\rm Mpc^{-1}} \, ,
\end{cases}
\label{eq:FDM-PS}
\end{eqnarray}
where $P_{\rm CDM}(k)$ is the power-spectrum for CDM cosmology.
The adopted FDM cosmology only depends on the particle mass $m_{\rm a}$, and the transfer function has a sharp cut-off, exhibiting acoustic oscillations on scales below $k_{\rm J,eq}$.
In this study, we adopt two different masses, $m_{\rm a} = 10^{-22}$\,eV (Run-A) and $10^{-21}$\,eV (Run-B) to investigate the effect on first star formation.
For comparison, we also carry out a simulation within standard $\Lambda$CDM cosmology (Run-Ref).
To control for cosmic variance, we generate all initial conditions with the same phases, assuming a Gaussian random process throughout.

The cosmological simulations are performed by using the parallel N-body/Smoothed Particle Hydrodynamics (SPH) code {\sc gadget-2} \citep{springel05b}, suitably modified for the primordial star formation case \citep{yoshida06,yoshida07,hirano14,hirano15a}.
In order to achieve the numerical resolution required to treat the wide dynamic range from cosmological scales to those of the gravitationally unstable gas cloud, we employ hierarchical zoom-in procedures.
Within a parent computational box of $10$\,comoving\,$h^{-1}$\,Mpc on a side,
we insert a series of nested refinement regions, reaching maximum resolution inside a volume with linear size $1~(0.5)$\,comoving\,$h^{-1}$\,Mpc for Run-A and B (Run-Ref).
The corresponding DM (N-body) and baryonic (SPH) particle masses in the maximally refined region are $119~(9.2)\,\msun$ and $21.9~(1.7)\,\msun$ for Run-A and B (Run-Ref), which can well resolve the halo structure.
We follow structure formation from redshift $z = 99$ until the maximum hydrogen number density reaches $10^6~\cc$.
During the simulations, we use an on-the-fly particle splitting technique \citep{kitsionas02}, with the refinement criterion that the Jeans mass is always resolved by at least $50$ SPH particles, which is sufficient to examine the hydrodynamics of fragmentation \citep{bate97}.
The most refined baryonic particle mass, as a result of splitting, thus reaches $1.7~(0.13)\,\msun$ for Run-A and B (Run-Ref), able to resolve the gravitationally unstable gas cloud.

In our simulations, we ignore the effect of FDM modified dynamics which causes the quantum interference patterns and solitonic cores in the DM distribution \citep[e.g.][]{woo09,schive14a,schive14b}.
The corresponding small-scale suppression may additionally delay the onset of star formation, but is likely sub-dominant compared with the overall delay due to the modified initial conditions.
Similarly, at the scale of the solitonic core the baryon density is already larger than the DM one, such that the subsequent hydrodynamics is largely decoupled from the DM distribution. The current uncertainty in the results is likely dominated by the unknown DM particle mass, compared to any uncertainties due to the neglected wave dynamics. Future, more complete simulations are needed to check these assumptions.

\begin{figure*}
\begin{tabular}{ccc}
\includegraphics[width=4.5cm]{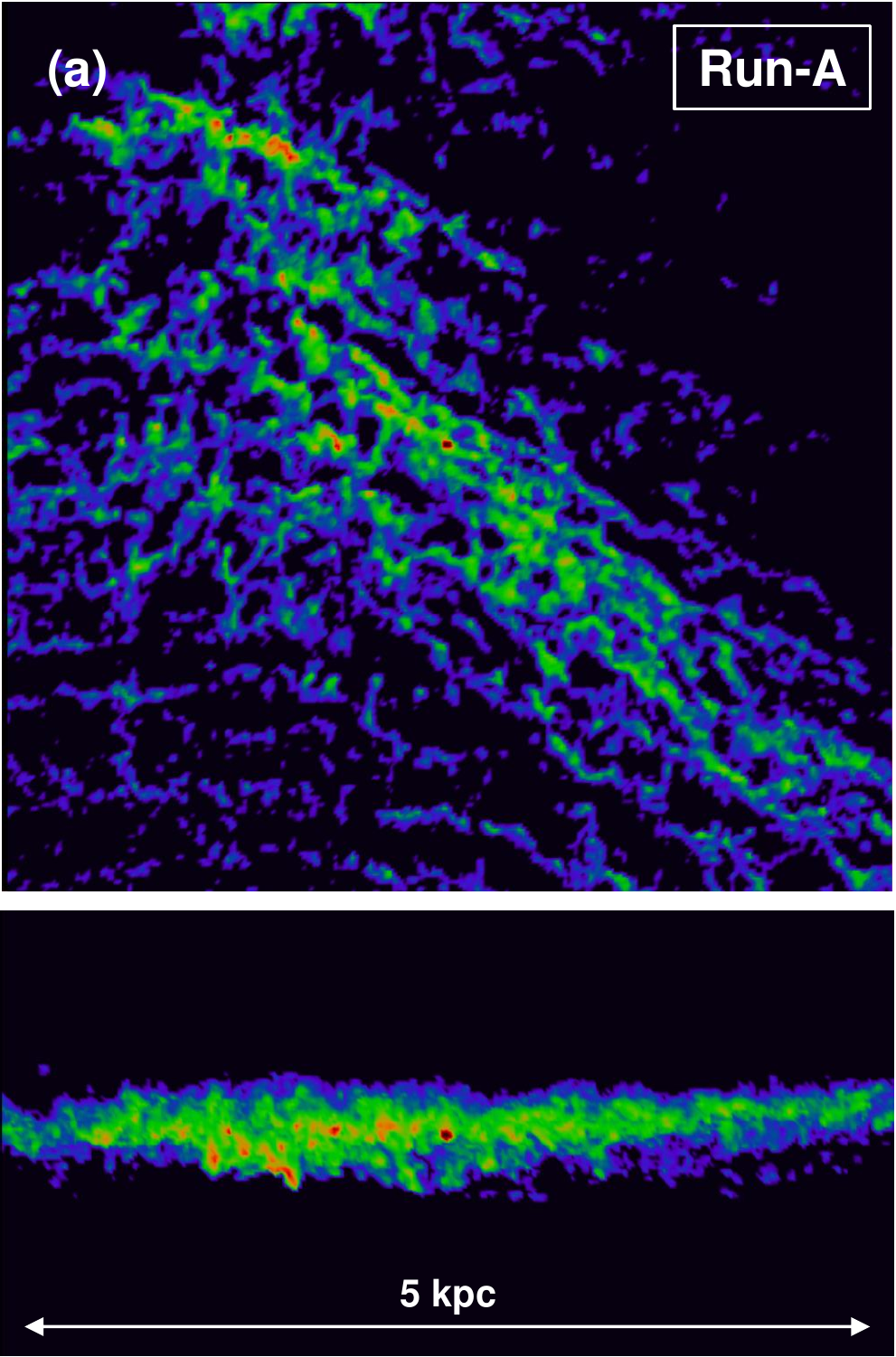}&
\includegraphics[width=4.5cm]{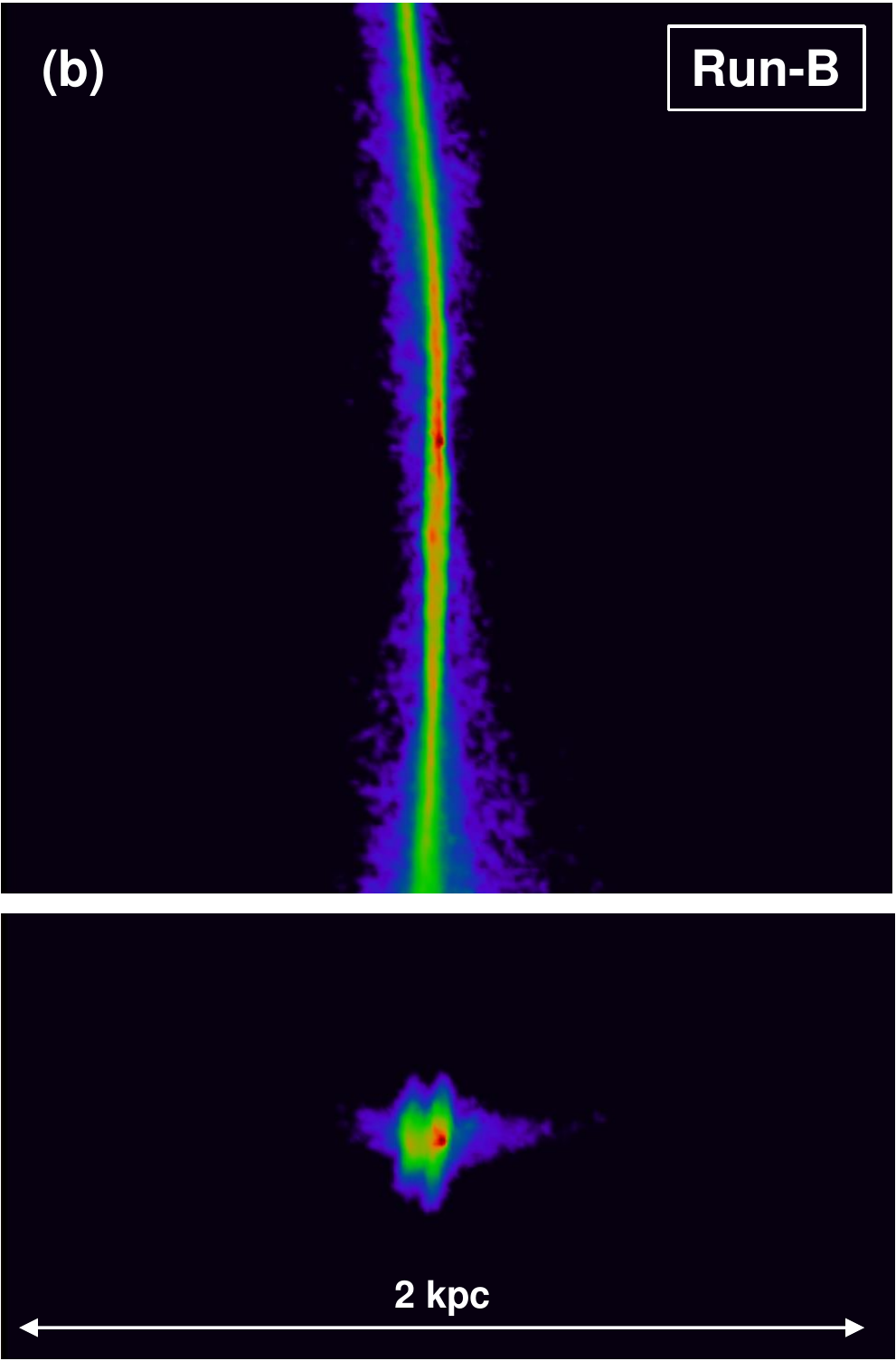}&
\includegraphics[width=4.5cm]{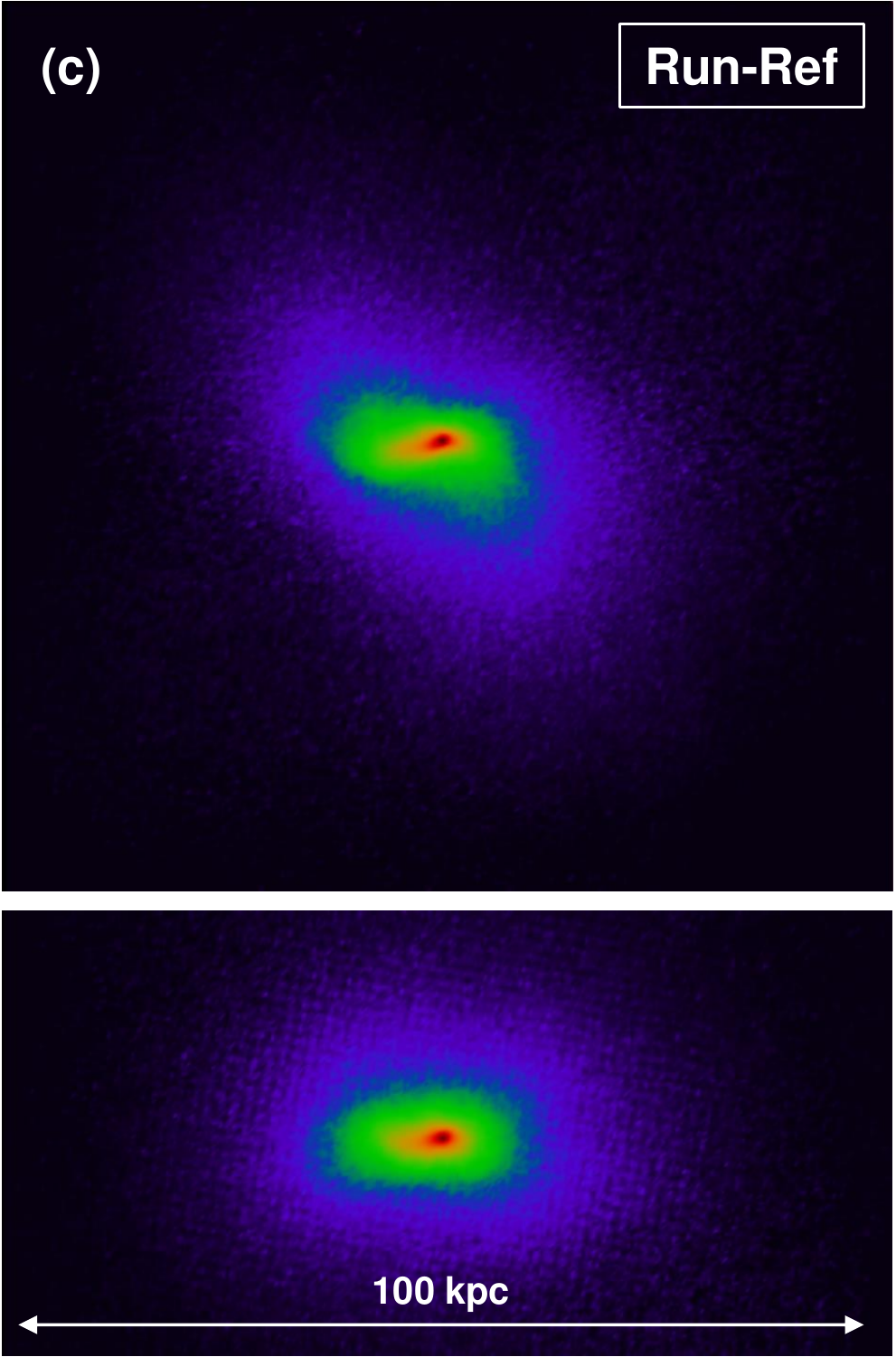}
\end{tabular}
\caption{
Morphology of primordial star formation in different cosmologies.
Cross-sectional views of gas density distribution around the collapse centre of the primordial star-forming cloud in Run-A with 5\,kpc linear size ({\it panel a}), Run-B with $2$\,kpc ({\it panel b}), and Run-Ref with $100$\,pc on the side ({\it panel c}).
The top and bottom panels show slices through the $X$-$Y$ and $X$-$Z$ planes, respectively.
The blue, green, orange, and red colours represent density contours, with $\nh = 1$, $10$, $10^2$, and $10^3\,\cc$, respectively.
The stark difference in morphology is evident, illustrating the shift from the near-spherical minihalo environment in $\Lambda$CDM to the filaments, or even sheets, of the cosmic web within FDM.
}
\label{f1}
\end{figure*}

\section{Star formation site}
\label{sec:halo}

The suppression of small-scale perturbations within FDM directly delays structure formation in the early Universe.
Star formation in the cosmological volume first occurs at $z = 6.5$ ($12.9$) in Run-A (B), which is significantly later than in Run-Ref at $z = 28.1$ (see Table~\ref{t1}).
We note that our virial mass,
given by the scale where the mean density is 200 times the background value,
and collapse redshift for Run-A with $m_{\rm a} \approx 10^{-22}$\,eV
are consistent with results from prior FDM structure formation simulations \citep{schive16}.
Such time delay results in a completely different morphology of the star formation site.
In $\Lambda$CDM cosmology, the first star formation events occur at the centre of almost spherically-symmetric dark matter minihaloes with $10^5$ -- $10^6\,\msun$ at $z \sim 20$ -- $30$ \citep[e.g.][]{hirano14}.
We compare the resulting star formation morphologies in Fig.~\ref{f1}, where the near-spherical geometry is evident in Run-Ref (Fig.~\ref{f1}c).
In the FDM cosmologies, on the other hand, minihaloes do not form, because the suppression scale of $\sim$\,kpc is larger than the minihalo scale in $\Lambda$CDM.
Pop~III stars, therefore, do not form in such spherical host regions, but instead in more massive DM sheets (Fig.~\ref{f1}a) or filaments (Fig.~\ref{f1}b)
whose virial masses are $10^3$ -- $10^4$ times the mass of a minihalo (Table~\ref{t1}).
In FDM models, the first objects in the Universe finally appear after $z \sim 15$, when super-kpc scale structures begin to collapse from large-scale perturbations.

\begin{table}
\begin{center}
\begin{tabular}{lrrrrr}
\hline
\hline
$m_{\rm a}$\,(eV) & $z$ & $R_{\rm vir}$\,(pc) & $M_{\rm vir}\,(\msun)$ & $T_{\rm vir}$\,(K) & $N_{\rm cloud}$ \\
\hline
$10^{-22}$ (A) &   6.5 & $4.5 \times 10^3$ & $3.9 \times 10^9$ & $8.0 \times 10^3$ & 10 \\
$10^{-21}$ (B) & 12.9 & $9.4 \times 10^2$ & $2.1 \times 10^8$ & $7.6 \times 10^3$ & a few \\
CDM (Ref)          & 28.1 & $4.9 \times 10^1$ & $5.6 \times 10^5$ & $1.9 \times 10^3$ & 1 \\
\hline
\end{tabular}
\caption{
Column 1: Mass of the ultra-light dark matter particle.
Column 2: Collapse redshift.
Column 3: Virial radius.
Column 4: Virial mass.
Column 5: Corresponding virial temperature.
Column 6: Estimated number of gravitationally unstable clumps.
Virial properties are calculated from the snapshots when the maximum gas density reaches $10^6\,\cc$.
}
\label{t1}
\end{center}
\end{table}

Generically, the objects found in the simulations emerge as components of the cosmic web, predicted by the standard model of structure formation \citep[e.g.][]{zeldovich70,shandarin89}.
A given perturbation on some scale in an almost uniform three-dimensional density distribution in the early Universe first collapses along one dimension, forming two-dimensional sheets.
Subsequently, these so-called `Zeldovich pancakes' collapse in another dimension and form one-dimensional filaments, where the order of collapse along the different axes is given by the eigenvalues of the deformation tensor.
Finally these filaments collapse in the remaining dimension and form three-dimensional spheres (or haloes).
The large suppression scale of the FDM particle prevents the later collapse at smaller scales, and our simulations demonstrate how first star formation consequently occurs during the initial phases of structure formation, inside the filaments, or even sheets, of the cosmic web.
Such a shift in primordial star formation site from haloes to filaments was already found for warm dark matter (WDM) models \citep{gao07b}, for similar reasons (discussed in Section~\ref{sec:summary}).

To further elucidate the difference between $\Lambda$CDM and FDM models,
Fig.~\ref{f2} shows the radial distribution of mass and density at the moment when the gas first undergoes runaway collapse.
The enclosed baryonic mass is significantly larger in the FDM runs compared to the $\Lambda$CDM one (Fig.~\ref{f2}a).
Within the virial radius, there are multiple density peaks in the FDM runs (Fig.~\ref{f2}b).
In Run-B, in particular, such secondary peaks are located along the dense filament (compare to Fig.~\ref{f1}b).
Such morphology, where multiple regions are becoming gravitationally unstable in a nearly synchronized fashion, is quite different from the $\Lambda$CDM case (Run-Ref).
Here, only one gravitationally unstable cloud forms at the centre of the almost spherical dark matter minihalo (Figs.~\ref{f1} and \ref{f2}b).

\section{Thermodynamics of collapsing cloud}
\label{sec:cloud}

The baryonic component is gravitationally dragged into the DM structure and finally begins self-gravitational collapse, when the gravitational force overcomes the thermal pressure.
This occurs when the mass of the growing gas cloud exceeds the local Jeans mass \citep{abel02}
\begin{eqnarray}
M_{\rm J} \approx 1000\,\msun\,\left( \frac{T}{200\,{\rm K}} \right)^{3/2} \left( \frac{\nh}{10^4\,\cc} \right)^{-1/2}\, ,
\label{eq:Mjeans}
\end{eqnarray}
where $T$ is the gas temperature and $\nh$ the hydrogen number density, normalized to the typical values encountered in primordial clouds.
The latter defines what is termed the `loitering state' of Pop~III star formation \citep{bromm02}.
The thermal evolution of the collapsing gas cloud governs the final fate of the new-born Pop~III protostars, because their mass accretion history, and thus the resulting stellar mass, sensitively depends on temperature as $\dot{M} \approx M_{\rm J} / t_{\rm ff} \propto T^{3/2}$, where $t_{\rm ff}$ is the free-fall time-scale.
In Fig.~\ref{f3}(a), we show the thermal histories of the collapsing clouds within the different DM models.
In the standard $\Lambda$CDM case (Run-Ref), the gas is adiabatically heated to the virial temperature of the minihalo \citep{barkana01},
\begin{eqnarray}
T_{\rm vir} \simeq 1500\,{\rm K}\,\left( \frac{M_{\rm vir}}{5 \times 10^5\,\msun} \right)^{2/3} \left( \frac{1+z}{30} \right) \, .
\label{eq:Tvir}
\end{eqnarray}
Subsequently, the cloud cools via H$_2$-cooling and finally becomes gravitationally unstable, when $M_{\rm enc} \ge M_{\rm J}$ at $\nh \sim 10^4\,\cc$, at the loitering state (see Fig.~\ref{f3}b).

\begin{figure}
\begin{center}
\includegraphics[width=0.7\columnwidth]{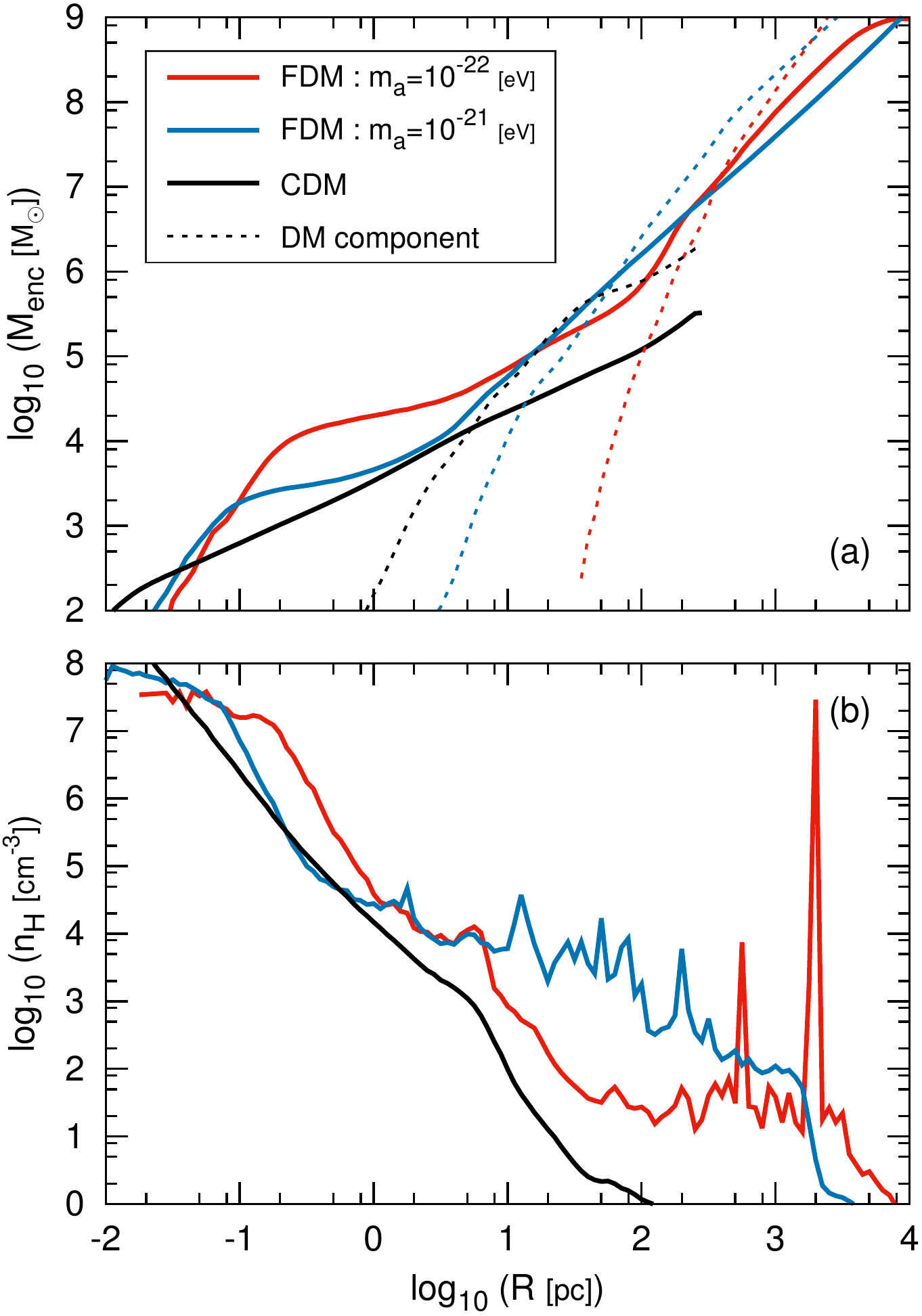}
\caption{
Radially averaged profiles of the enclosed gas mass ({\it panel a}) and density ({\it panel b}), at the moment when the maximum density reaches $10^6\,\cc$.
The red, blue, and black lines represent results at the end of the simulation for Run-A, Run-B, and Run-Ref, respectively.
The solid and dashed lines in panel (a) show baryonic and DM components, respectively.
The structural differences between the models are readily apparent, in particular regarding the more distributed nature of gravitational instability in the FDM models.
}
\label{f2}
\end{center}
\end{figure}

\begin{figure}
\begin{center}
\includegraphics[width=0.7\columnwidth]{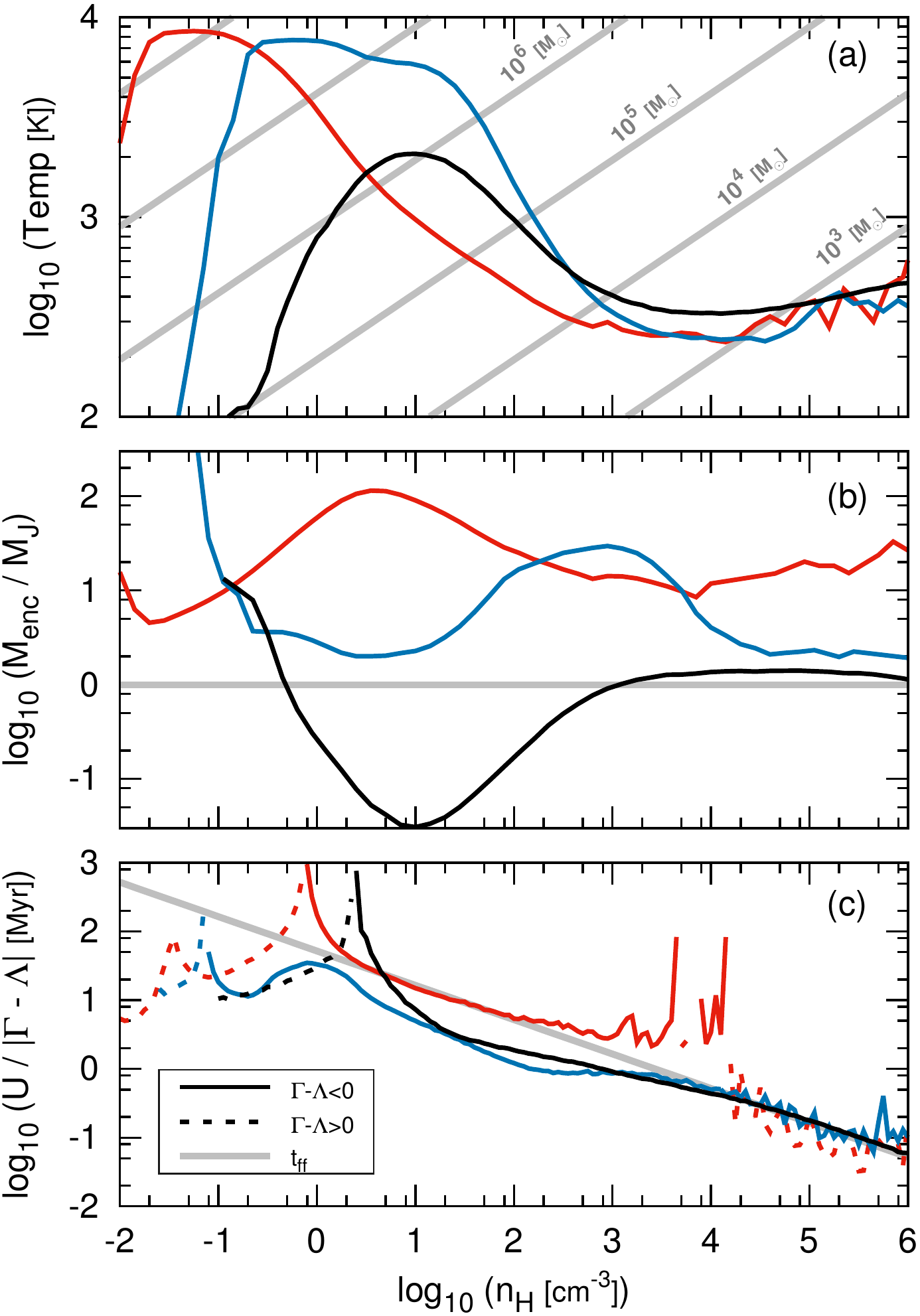}
\caption{
Profiles of collapsing cloud properties as a function of hydrogen number density: gas temperature ({\it panel a}), ratio of enclosed to Jeans mass ({\it panel b}), and thermal time-scale where $U$, $\Gamma$, and $\Lambda$ are the internal energy, heating rate, and cooling rate, respectively ({\it panel c}).
For the red, blue, and black lines, we adopt the same convention as in Fig.~\ref{f2}.
The grey lines in panel (a) show the density-temperature relation for a given Jeans mass (Eq.~\ref{eq:Mjeans}; $10^3$ -- $10^8\,\msun$).
The grey line in panel (c) represents the corresponding free-fall time-scale as a function of density, $t_{\rm ff} = \sqrt{3 \pi /32 G \rho} = 5.2 \times 10^7 (\nh/\cc)^{-1/2}$\,yr.
It is evident that the thermal histories of the primordial gas approach a very similar behaviour at high densities, implying similar fragmentation properties ({\it panel a}).
The `Jeans number' $M_{\rm enc}/M_{\rm J}$, on the other hand, is greatly enhanced in the FDM cases, implying a possibly larger star formation efficiency in turn.
}
\label{f3}
\end{center}
\end{figure}

Within FDM, the first structures to virialize are more massive compared to $\Lambda$CDM, such that the primordial gas collapses into deeper potential wells with correspondingly larger virial temperatures.
During the collapse, the gas temperature initially increases through adiabatic compression, until it reaches $T \sim 8$,$000$\,K, when atomic hydrogen (H) line cooling becomes efficient in the partially ionized gas.
The strong atomic cooling balances any further gravitational heating, such that the gas clouds never reach the virial temperature of their DM host structure.
Specifically the H-cooling temperature plateau of $8$,$000$\,K is less than the corresponding virial value as a function of virial mass and redshift (Eq.~\ref{eq:Tvir}), where $1.3 \times 10^5$\,K for Run-A, and $3.5 \times 10^4$\,K for Run-B.
The massive gas clouds can be supported until H$_2$-cooling becomes effective (Fig.~\ref{f3}c), when the gas temperature finally declines, and the Jeans mass decreases (Eq.~\ref{eq:Mjeans}).

The massive gas clouds with $10^7$ -- $10^8\,\msun$ are initially heated up to $8$,$000$\,K,
but finally cool to $200$\,K, which is below the minimum temperature in Run-Ref, via enhanced H$_2$-cooling.
This boost in cooling results from the increased free electron fraction which in turn catalyses additional H$_2$ formation.
In such low-temperature and H$_2$-rich gas, additional cooling via the HD molecule could become efficient and cool the gas even further \citep[e.g.][]{nakamura02,gao07b}.
In the FDM cases studied here, however, the HD fraction remains below the critical value required for cooling to affect the cloud's thermal evolution.
This is because the gravitationally unstable gas clouds (Fig.~\ref{f3}b) collapse too rapidly for sufficient HD molecule formation \citep[see fig.~23 in][]{hirano14}.

Due to the delayed collapse and the corresponding increase in virial mass scale, more massive gas clouds become gravitationally unstable in the FDM universe.
The temperatures encountered by those clouds, on the other hand, are similar to the standard $\Lambda$CDM case (Run-Ref), such that the corresponding Jeans masses approach similar values as well.
In Run-Ref, the mass of the collapsing cloud first exceeds the local Jeans mass with about $10^3\,\msun$ at the loitering point, whereas a gas cloud with more than $10^5\,\msun$ becomes Jeans unstable during the temperature decline phase in Run-A and B.
These differences may result in a qualitatively altered star formation process, where star formation efficiencies could be much higher, and a massive cluster of Pop~III stars could arise.
Such a primordial starburst would be markedly different from the standard prediction for minihaloes, where small groups of Pop~III stars, including binaries and small multiples, are expected to form \citep[e.g.][]{bromm13}.
As a simple estimation of the number of gravitationally-unstable clumps in the FDM runs, we could consider that the $10^5\,\msun$ cloud fragments at the final Jeans scale of $10^3\,\msun$, resulting in about 100 star-forming clumps.
Slightly more precise, we can measure the mass of the gas which actually reaches the minimum temperature, giving about ten (a few) times the Jeans mass in Run-A (Run-B).
In fact, a number of independent clumps can be discerned in the density distribution (Figs.~\ref{f1} and \ref{f2}).

\section{Discussion and summary}
\label{sec:summary}

We present a series of cosmological simulations with ultra-light particle dark matter, suggesting an entirely different process of first star formation in the early Universe.
Multiple star-forming clouds, formed within sheet-like or filamentary massive structures, become gravitationally unstable at the same time.
Although their detailed subsequent evolution is beyond the scope of the current work, the further fate of such primordial proto-clusters is clearly of great interest in terms of the effect on the chemical and radiative evolution at early epochs, possibly shaping the assembly of the first galaxies and the reionization process.
Standard theory posits that massive star clusters can only form in metal-enriched gas \citep[e.g.][]{safranek14}. Intriguingly, within FDM cosmologies, such massive starbursts could already occur at the very onset of cosmic star formation.
The extreme time delay of the first objects in the FDM vs. $\Lambda$CDM models, with collapse occurring at $z = 6.5$ vs. $28.1$ for our cases, can provide a sensitive astrophysical probe to constrain the nature of the dark matter particle.
Upcoming deep-field observations with the {\it JWST}, and beyond that  with the suite of extremely large telescopes on the ground (the GMT, TMT, and E-ELT), will conduct searches for the timing and intensity of such primordial (Pop~III) starbursts, as predicted for FDM cosmologies.

There are alternative origins which can reproduce similar environmental conditions for massively-clustered first star formation.
One such scenario is adopting warm dark matter (WDM) particles with masses of $\approx\!1$\,keV.
Cosmological simulations in the WDM Universe have been performed to investigate its effect on the formation process of the first stars \citep{gao07b}, showing the formation of very massive filaments with $10^7\,\msun$.
\citet{gao07b} also discuss the possibility of filament fragmentation and subsequent coalescence.
We find similar filamentary structure in Run-B, and also multiple gravitational collapse events along the filament.
Furthermore, in Run-A for the lower-mass axion, a sheet-like structure initially forms and fragments into multiple clumps.
Additional, qualitatively very different physical origins exist to realize a similar thermal history of the primordial gas.
Among them are the collapse within standard $\Lambda$CDM, but exposed to a strong external soft ultra-violet radiation field which photo-dissociates the H$_2$ molecule until the gas cloud becomes sufficiently dense to self-shield \citep[e.g.][]{latif15}, or gas collapse subject to the impeding effect of relative baryon-DM streaming \citep[e.g.][]{tseliakhovich10,latif14,schauer17}.
In these cases, however, the small-scale DM perturbations still exist, leading to dynamical differences to the FDM (or WDM) models.

In the possible post-WIMP era of dark matter physics, astrophysical constraints will have to play a unique role in guiding particle physics theory.
This is in particular the case for high-redshift structure formation, encompassing the epochs of first star formation and reionization, as there we encounter the small-scale consequences of any DM models which have hitherto been largely
beyond reach for empirical testing.
The imminent arrival of the next generation of telescopes may thus herald exciting new discoveries not only for astronomy, but also for fundamental physics.

\section*{Acknowledgements}

The simulations were carried out on
XC30 at CfCA, National Astronomical Observatory of Japan,
XC40 at YITP, Kyoto University, and
Stampede at TACC, University of Texas at Austin.
This work was financially supported by
Grant-in-Aid for JSPS Overseas Research Fellowships (S.H.) and
NSF grant AST-1413501 (V.B.).

\bibliographystyle{mnras}
\bibliography{biblio}

\bsp
\label{lastpage}
\end{document}